\documentclass{svproc}

\usepackage[utf8]{inputenc}
\usepackage[toc,page]{appendix}
\usepackage[utf8]{inputenc}
\usepackage{graphicx}
\usepackage[hscale=0.7,vscale=0.8]{geometry}
\usepackage{amsmath}
\usepackage{amssymb}
\usepackage{verbatim}
\usepackage{soul}
\usepackage{url}
\usepackage{xcolor}
\usepackage{textcomp}
\usepackage{multicol}

\usepackage{blindtext}
\usepackage{academicons}
\usepackage{authblk}
\usepackage{babel}

\newcommand{\orcid}[1]{{#1}}

\title{Experiences and Lessons Learned\\ 
Creating and Validating Concept Inventories\\ for Cybersecurity}

\author{Alan T. Sherman\inst{1}\orcid{0000-0003-1130-4678}
\and Geoffrey L. Herman\inst{2}\orcid{0000-0002-9501-2295}
\and \\Linda Oliva\inst{1}\orcid{0000-0003-0056-7819}
\and Peter A. H. Peterson\inst{3}\orcid{0000-0003-3285-496X}
\and \\Enis Golaszewski\inst{1}\orcid{0000-0002-0814-9956}
\and Seth Poulsen\inst{2}\orcid{0000-0001-6284-9972}
\and \\Travis Scheponik\inst{1}\orcid{0000-0002-0821-1123}
\and Akshita Gorti\inst{1}\orcid{0000-0003-2208-165X}
}
\authorrunning{Alan T. Sherman et al.}

\institute{Cyber Defense Lab, University of Maryland, Baltimore County (UMBC), Baltimore, MD 21250,
\email{sherman@umbc.edu},
\url{www.csee.umbc.edu/people/faculty/alan-t-sherman/}
\and
Computer Science, University of Illinois at Urbana-Champaign, Champaign, IL 61820,
\email{glherman@illinois.edu},
\url{publish.illinois.edu/glherman/}
\and
Department of Computer Science, University of Minnesota Duluth, Duluth, MN 55812,
\email{pahp@d.umn.edu},
\url{www.d.umn.edu/~pahp/}
}

\date{{\today} (preliminary draft)}

\begin{document}

\maketitle

\begin{abstract}

 We reflect on our ongoing journey in the 
 educational {\it Cybersecurity Assessment Tools (CATS)} Project
 to create two concept inventories for cybersecurity.
 We identify key steps in this journey and important questions we faced.  
 We explain the decisions we made and discuss the consequences of those decisions, 
 highlighting what worked well and what might have gone better.

\hspace*{0.2in}
 The CATS Project is creating and validating two 
 concept inventories---conceptual tests of understanding---that can be used
 to measure the effectiveness of various approaches to teaching and learning cybersecurity.
 The {\it Cybersecurity Concept Inventory (CCI)} is for students 
 who have recently completed any first course in cybersecurity;
 the {\it Cybersecurity Curriculum Assessment (CCA)} is for students 
 who have recently completed an undergraduate major or track in cybersecurity.
 Each assessment tool comprises 25 {\it multiple-choice questions (MCQs)}
 of various difficulties
 that target the same five core concepts, but the CCA assumes greater
 technical background.
 
 \hspace*{0.2in}
 Key steps include 
 defining project scope, 
 identifying the core concepts, 
 uncovering student misconceptions, 
 creating scenarios, 
 drafting question stems, 
 developing distractor answer choices, 
 generating educational materials,
 performing expert reviews, 
 recruiting student subjects, 
 organizing workshops, 
 building community acceptance, 
 forming a team and nurturing collaboration, 
 adopting tools, and 
 obtaining and using funding.

\hspace*{0.2in}
 Creating effective MCQs is difficult and time-consuming, and cybersecurity presents
 special challenges.  Because cybersecurity issues are often subtle,
 where the adversarial model and details matter greatly, it is challenging 
 to construct MCQs for which there is exactly one best but non-obvious answer.
 We hope that our experiences and lessons learned may help others 
 create more effective concept inventories and assessments in STEM.
 
\keywords{
Computer science education \middot
Concept inventories \middot
Cryptography \middot
Cybersecurity Assessment Tools (CATS) \middot
Cybersecurity education \middot
Cybersecurity Concept Inventory (CCI) \middot
Cybersecurity Curriculum Assessment (CCA) \middot
Multiple-choice questions.}

\end{abstract}

\clearpage
\section{Introduction}
\label{sec:introduction}

When we started the {\it Cybersecurity Assessment Tools (CATS)} Project~\cite{CATS2017} in 2014, we thought that it should not be difficult to create a collection of 
25 {\it multiple choice questions (MCQs)} that assess student understanding of core cybersecurity concepts.  Six years later, in the middle of validating the two draft assessments we have produced, we now have a much greater appreciation for the significant difficulty of creating and validating effective and well-adopted concept inventories.  This paper highlights and reflects on critical steps in our journey, with the hope that our experiences can provide useful lessons learned to anyone who wishes to create a cybersecurity concept inventory, any assessment in cybersecurity, or any assessment in STEM.\footnote{Science, Technology, Engineering, and Mathematics (STEM).}

Cybersecurity is a vital area of growing importance for national competitiveness, and 
there is a significant need for cybersecurity professionals~\cite{BOL2020}. 
The number of cybersecurity programs 
at colleges, universities, and training centers is increasing.
As educators wrestle with this demand, there is a corresponding awareness that we lack a rigorous research base that informs how to prepare cybersecurity professionals. 
Existing certification exams, such as CISSP~\cite{cissp}, 
are largely informational, not conceptual. 
We are not aware of any scientific analysis of any of these exams. 
Validated assessment tools are essential so that cybersecurity educators have trusted methods for discerning whether efforts to improve student preparation are successful~\cite{Olds2005}.  
The CATS Project provides rigorous evidence-based instruments for assessing and evaluating educational practices;  in particular, they will help assess approaches to teaching and learning cybersecurity such as
traditional lecture, case study, hands-on lab exercises, interactive simulation, competition, and gaming.



We have produced two draft assessments, each comprising 25 MCQs.  The {\it Cybersecurity Concept Inventory (CCI)} measures how well students understand
core concepts in cybersecurity after a first course in the
field. The {\it Cybersecurity Curriculum Assessment (CCA)} measures how
well students understand core concepts after completing a full cybersecurity curriculum.
Each test item comprises a {\it scenario}, a {\it stem} (a question), and five {\it alternatives} (answer choices comprising a single best answer choice and four distractors).
The CCI and CCA target the same five core concepts (each being an aspect of adversarial thinking), but the CCA assumes greater technical background.  In each assessment, there are five test items of various difficulties for each of the five core concepts.
   
Since fall 2014, we have been following prescriptions of the National Research Council for developing rigorous and valid assessment tools~\cite{Jorion2015,Pel01}. We carried out two surveys using the Delphi method to identify the scope and content of the assessments~\cite{Delphi2016}. 
Following~\cite{Ericsson:1993}, we then carried out qualitative interviews~\cite{Scheponik2016,Thompson2018} to develop a cognitive theory that can guide the construction of assessment questions. Based on these interviews, we have developed a preliminary battery of over 30 test items for each assessment. Each test item measures how well students understand core concepts as identified by our Delphi studies. The distractors (incorrect answers) for each test item are based in part on student misconceptions observed during the interviews.  We are now validating the CCI and CCA using small-scale pilot testing, cognitive interviews, expert review, and large-scale psychometric testing~\cite{FIE2019}.

The main contributions of this paper are lessons learned from our experiences with the CATS Project. These lessons include strategies 
for developing effective scenarios, stems, and distractors, recruiting subjects for psychometric testing, and building and nurturing effective collaborations.  We offer these lessons, not with the intent of prescribing advice for all, but with the hope that others may benefit through understanding and learning from our experiences. This paper aims to be the paper we wished we could have read before starting our project.

\section{Background and Previous and Related Work}
\label{sec:background}

We briefly review relevant background on concept inventories, 
cybersecurity assessments, and other related work.
To our knowledge, our CCI and CCA are the first concept inventories for cybersecurity, 
and there is no previous paper that presents lessons learned
creating and validating any concept inventory.

\subsection{Concept Inventories}
\label{sec:CIs}

A {\it concept inventory (CI)} is an assessment (typically multiple-choice)
that measures how well student conceptual knowledge aligns with accepted conceptual 
knowledge~\cite{Hes92}. 
Concept inventories have been developed for many STEM disciplines, consistently revealing that students who succeed on traditional classroom assessments struggle to answer deeper conceptual 
questions~\cite{Chi2006,Evans2003,Hake1998,Hes92,Litzinger2010}. 
When students have accurate, deep conceptual knowledge, they can learn more efficiently, and they can transfer their knowledge across 
contexts~\cite{Litzinger2010}.
CIs have provided critical evidence supporting the adoption of active learning and other evidence-based 
practices~\cite{Evans2003,Hake1998,Hake2001,Mestre93}.
For example, the {\it Force Concept Inventory (FCI)} by Hestenes et al.~\cite{Hes92} ``spawned a dramatic movement of reform in physics education.''~\cite[p.\ 1018]{Epstein2013}.


For CIs to be effective, they need to be validated. Unfortunately, few CIs have undergone rigorous 
validation~\cite{Pellegrino2011,Wallace2010}. 
Validation is a chain of evidence that supports the claims that an assessment measures the attributes that it claims to measure. This process requires careful selection of what knowledge should be measured, carefully constructing questions that are broadly accepted as measuring that knowledge, and providing statistical evidence that the assessment is internally consistent. The usefulness of a CI is threatened if it fails any of these requirements. Additionally, a CI must be easy to administer, and its results must be easy to interpret---or they can easily be misused. Critically, CIs are intended as research instruments that help instructors make evidence-based decisions to improve their teaching and generally should not be used primarily to assign student grades or to evaluate a teacher's effectiveness.

Few validated CIs have been developed for computing topics; notable exceptions include the Digital Logic Concept Inventory~\cite{Herman2010} (led by CATS team member Herman) and 
early work on the Basic Data Structures Inventory~\cite{Porter2019}. 
None has been developed for security related topics.

\subsection{Cybersecurity Assessment Exams}
\label{sec:cyberexams}

We are not aware of any other group that is developing an educational assessment tool for cybersecurity.  There are several existing certification exams, including ones
listed by NICCS as relevant~\cite{NICCS}.


CASP+~\cite{casp} comprises multiple-choice and performance tasks items including enterprise security, risk management, and incident response. 
OSCP~\cite{oscp} (offensive security) is a 24-hour practical test focusing on penetration testing. 
Other exams include CISSP, Security+, and
CEH~\cite{security+,cissp,ceh},
which are mostly informational, not conceptual.  
Global Information Assurance Certification
(GIAC)~\cite{giac}
offers a variety of vendor-neutral MCQ certification exams linked to SANS courses;
for each exam type, the gold level requires a research paper.
We are unaware of any scientific study that characterizes the properties of any of these tests.

\subsection{Other Related Work} 
\label{sec:relatedwork}

The 2013 IEEE/ACM Computing Curriculum Review~\cite{ACM13} approached the analysis of cybersecurity content in undergraduate education from the perspective of traditional university curriculum development. Later, the ACM/IEEE/AIS SIGSEC/IFIP Joint Task Force on Cybersecurity Education (JTF)~\cite{CSEC} developed comprehensive curricular guidance in cybersecurity education, releasing Version~1.0 of their guidelines at the end of 2017.


To improve cybersecurity education and research, the National Security Agency (NSA) and 
Department of Homeland Security (DHS) jointly sponsor the 
National Centers of Academic Excellence (CAE) program. 
Since 1998, more than 300 schools have been designated as CAEs in Cyber Defense.
The requirements include sufficiently
covering certain ``Knowledge Units'' (KUs)
in their academic programs, making the CAE program a 
``significant influence on the curricula of programs offering cybersecurity education''~\cite{Gibson19}. 

The NICE Cybersecurity Workforce Framework~\cite{NICE} establishes a common lexicon for explaining a structured description of professional cybersecurity positions in the workforce with detailed documentation of the knowledge, skills, and abilities needed for various types of cybersecurity activities. 

More recently, the Accreditation Board of Engineering and Technology (ABET) has included, in the 2019--2020 Criteria for Accrediting Computing Program, criteria for undergraduate cybersecurity (or similarly named) programs. ABET has taken an approach similar to that of the CAE program, requiring coverage of a set of topics without requiring any specific set of courses.

In a separate project, CATS team member Peterson and his students~\cite{Jindeel19,peter-mis} worked with experts to identify specific and persistent commonsense misconceptions in cybersecurity, such as that ``physical security is not important,'' or that ``encryption is a foolproof security solution.'' They are developing a CI focusing on those misconceptions.

\section{Key Steps and Takeaways from the CATS Project}
\label{sec:experiences}

We identify the key steps in our journey creating and validating 
the CCI and CCA.  For each step, we comment on important issues, 
decisions we made, the consequences of those decisions,
and the lessons we learned.

\subsection{Genesis of the CATS Project} 
\label{sec:genesis}

On February 24--25, 2014, Sherman, an expert in cybersecurity, participated in a NSF workshop to advise NSF on how to advance cybersecurity education. 
NSF occasionally holds such workshops in various areas and distributes their reports, which can be very useful in choosing research projects.
The workshop produced a list of prioritized recommendations, beginning with the creation of a concept inventory~\cite{GWU14}. At the workshop, Sherman met one of his former MIT officemates, Michael Loui. Sherman proposed to Loui that they work together
to create such a concept inventory. About to retire, Loui declined, and introduced Sherman to Loui's recent PhD graduate Herman, an expert in engineering education. Without meeting in person for over a year, Sherman and Herman began a productive collaboration. Loui's introduction helped establish initial mutual trust between Sherman and Herman.

\subsection{Defining Scope of Project and Assessment Tools} 
\label{sec:scope}

As with many projects, defining scope was one of the most critical decisions
of the CATS Project.
We pondered the following questions, each of whose answers had profound implications on the
direction and difficulty of the project.
How many assessment tools should we develop? 
For what purposes and subject populations should they be developed?  
In what domain should the test items be cast?  
Should the test items be MCQs?  

We decided on creating two tools: the CCI (for students in any first course in cybersecurity) and 
CCA (for recent graduates of a major or track in cybersecurity), because there is a strong need for each, and each tool has different requirements. This decision doubled our work.
Creating any more tools would have been too much work.

Our driving purpose is to measure the effectiveness of various approaches to teaching cybersecurity,
not to evaluate the student or the instructor.  This purpose removes the need for a high-stakes test that would require substantial security and new questions for each test instance.  By contrast, many employers who have talked with us about our work have stated their desire for an instrument that would help
them select whom to hire (our assessments are neither designed nor validated for that high-stakes purpose).

Ultimately, we decided that all test items should be cast in the domain of cybersystems, on the grounds that cybersecurity takes place in the context of such systems.  Initially, however, we experimented with developing test items that probed security concepts more generally, setting them in a variety of 
every-day contexts, such as building security, transportation security, and physical mail.
Both approaches have merit but serve different purposes.


Following the format of most concept inventories, we decided that each test item be a MCQ.
For more about MCQs and our reasons for using them, see Section~\ref{sec:discussion}.

\subsection{Identifying Core Concepts} 
\label{concepts}

The first major step in creating any concept inventory is to identify the core concepts to be tested. We sought about five important, difficult,  timeless, cross-cutting concepts.  These concepts do not have to cover cybersecurity comprehensively. For example, the Force Concept Inventory targets five concepts from Newtonian dynamics, not all concepts from physics.
To this end, we engaged 36 cybersecurity experts in two Delphi processes, one for the CCI and one for the CCA~\cite{Delphi2016}. A Delphi process is a structured process for achieving consensus on contentious issues~\cite{Brown1968,Gol2010}.  

An alternative to the Delphi process is the focus group.  Although focus groups can stimulate discussions, they can be influenced strongly by personalities and it can be difficult to organize the results coherently.  For example, attempts to create concept maps for cybersecurity via focus groups have struggled to find useful meaning in the resulting complex maps, due to their high density.\footnote{Personal correspondence with Melissa Dark (Purdue).}

Delphi processes also have their challenges, including recruiting and retaining experts, 
keeping the experts focused on the mission, and processing expert comments, including
appropriately grouping similar comments.  We started with 
36 experts in total, 33 for CCI, 31 for CCA, and 29 in both.
We communicated with the experts via email and SurveyMonkey.
For each process, approximately 20 experts sustained their efforts throughout.
Many of the experts came with strongly held opinions to include their favorite topics, such as
policy, forensics, malware analysis, and economic and legal aspects.  
We completed the two Delphi processes in parallel in fall 2014, taking about eight weeks, conducting initial topic identification followed by three rounds of topic ratings.  Graduate research assistant Parekh helped orchestrate the processes.
It is difficult to recruit and retain experts, and it is a lot of work to process the large
volume of free-form comments.

The first round produced very similar results for both Delphi processes, with both groups strongly identifying aspects of adversarial thinking.  Therefore, we restarted the CCI process with an explicit focus on adversarial thinking.  
After each round, using principles of grounded theory~\cite{Glaser1968},
we grouped similar responses and asked each expert to rate each response on a scale
from one to ten for importance and timeliness. We also encouraged experts to explain their
ratings.  We communicated these ratings and comments (without attribution) to everyone.
The CCA process produced a long list of topics, with the highest-rated ones embodying aspects of adversarial thinking.

In the end, the experts came to a consensus on five important core concepts, 
which deal with adversarial reasoning (see Table~\ref{tab:concepts}).
We decided that each of the two assessment tools would target these same five concepts, but assume different levels of technical depth.

\stepcounter{footnote}
\begin{table}
\begin{center}
  \caption{The five core concepts underlying the CCI and CCA embody aspects of adversarial thinking.}
  \begin{tabular}{ll}
  1 & Identify vulnerabilities and failures\\
  2 & Identify attacks against CIA triad$^{\arabic{footnote}}$ and authentication\\
  3 & Devise a defense\\
  4 & Identify the security goals\\
  5 & Identify potential targets and attackers\\
  \end{tabular}
  \label{tab:concepts}
\end{center}
\end{table}

\footnotetext{CIA Triad (Confidentiality, Integrity, Availability).}

\vspace*{-32pt}  

\subsection{Interviewing Students} 
\label{sec:interviews}

We conducted two types of student interviews: talk-aloud interviews to uncover student misconceptions~\cite{Thompson2018}, and cognitive interviews as part of the validation process~\cite{FIE2019}. We conducted the interviews with students from three diverse schools:
UMBC (a public research university), Prince George's Community College, and 
Bowie University (a Historically Black College or University (HBCU)).
UMBC's Institutional Review Board (IRB) approved the protocol.

We developed a series of scenarios based on the five core concepts identified in the Delphi processes.  Before drafting complete test items, we conducted 26 one-hour talk-aloud interviews to uncover misconceptions, which we subsequently used to generate distractors.  During the interviews we asked open-ended questions of various difficulties based on prepared scenarios.  For each scenario, we also prepared a  ``tree'' of possible hints and follow-up questions, based on the student's progress.  The interviewer explained that they wanted to understand how the student thought about the problems, pointing out that the interviewer was not an expert in cybersecurity and that they were not evaluating the student.  One or two cybersecurity experts listened to each interview, but reserved any possible comments or questions until the end.  We video- and audio-recorded each interview.

We transcribed each interview and analyzed it using novice-led 
paired thematic analysis~\cite{Thompson2018}.  Labeling each section of each interview 
as either ``correct'' or ``incorrect,'' 
we analyzed the data for patterns of misconceptions.
Four themes emerged: overgeneralizations, conflated concepts, biases, and incorrect assumptions~\cite{Thompson2018}. 
Together, these themes reveal that students generally failed to grasp the complexity and subtlety of possible vulnerabilities, threats, risks, and mitigations.

As part of our validation studies, we engaged students in cognitive interviews during which a student reasoned aloud while they took the CCI or CCA. These interviews helped us determine if students understood the questions, if they selected the correct answer for the correct reason, and if they selected incorrect answers for reasons we had expected.  These interviews had limited contributions since most subjects had difficulty providing  rationales for their answer choices.  The interviews did reveal that specific subjects had difficulty with some of the vocabulary, prompting us to define
selected terms (e.g., masquerade) at the bottom of certain test items.


Although there is significant value in conducting these interviews, they are a lot of work, especially analysis of the talk-aloud interviews.  For the purpose of generating distractors, we now recommend very strongly the simpler technique of asking students (including through crowdsourcing) open-ended stems, without providing any alternatives (see Section~\ref{sec:distractors}). 

\subsection{Creating Scenarios} 
\label{sec:scenarios}

To prepare for our initial set of interviews (to uncover student misconceptions), we created several interview prompts, each based on an engaging scenario.  Initially we created twelve scenarios organized in three sets of four, each set including a variety of settings and difficulty levels. 

We based our first CCI test items on the initial twelve scenarios, each test item comprising a scenario, stem, and five answer choices.  Whenever possible, to keep the stem as simple as possible, we placed details in the scenario rather than in the stem.  Initially, we had planned to create several stems for each scenario, but as we explain in Section~\ref{sec:discussion}, often this plan was hard to achieve.  Over time, we created many more scenarios, often drawing from our life experiences.  Sometimes we would create a scenario specifically intended to target a specific concept (e.g., identify the attacker) or topic (e.g., cyberphysical system).

For example, one of our favorite CCI scenarios is a deceptively simple one based on lost luggage.
We created this scenario to explore the concept of identifying targets and attackers.

\begin{quotation} 
\noindent {\bf Lost Luggage.} \emph{Bob's manager Alice is traveling abroad to give a sales presentation about an important new product. Bob receives an email with the following message: ``Bob, I just arrived and the airline lost my luggage. Would you please send me the technical specifications? Thanks, Alice.''}
\end{quotation}

Student responses revealed a dramatic range of awareness and understanding of core cybersecurity concepts. Some students demonstrated lack of adversarial thinking in suggesting that Bob should simply e-mail the information to Alice, reflecting lack of awareness of potential threats, such as someone impersonating Alice or eavesdropping on the e-mail. Similarly, others recognized the need to authenticate Alice, but still recommended e-mailing the information without encryption after authenticating Alice.  A few students gave detailed thoughtful answers that addressed a variety of concerns including authentication, confidentiality, integrity, policy, education, usability, and best practices.

\begin{figure}[t] 

\begin{quotation} 
\noindent {\bf Switchbox.}
\it A company has two internal Local Area Networks (LANs): a core LAN connected to an email server and the Internet, and an accounting LAN connected to the corporate accounting server (which is not connected to the Internet). Each desktop computer has one network interface card. Some computers are connected to only one of the networks (e.g., Computers~A and~C).  A computer that requires access to both LANs (e.g., Computer~B) is connected to a switchbox with a toggle switch that physically connects the computer to exactly one LAN at a time.
\end{quotation}

    \centering
    \includegraphics[width=0.75\textwidth]{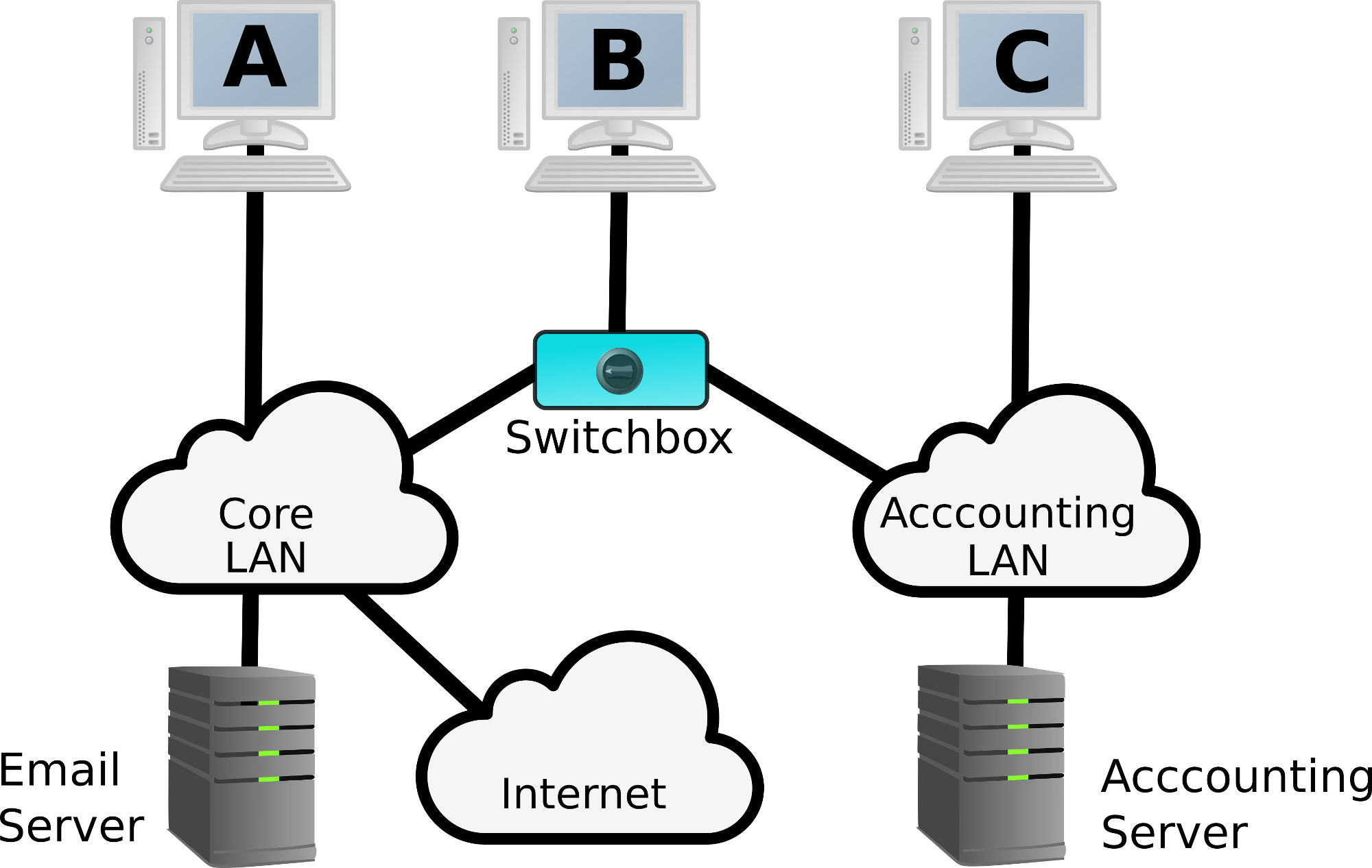}
    \caption{CCA switchbox scenario, which includes an artifact of a network diagram with switchbox. }
    \label{fig:switchbox}
\end{figure}

We designed the CCA for subjects with greater technical sophistication, 
for which scenarios often include an artifact (e.g., program, protocol, log file, system diagram, or product specification). 
We based some CCA test items directly on CCI items, adding an artifact.  In most cases we created entirely new scenarios.
In comparison with most CCI scenarios, CCA scenarios with artifacts require students to reason about more complex real-world challenges in context with specific technical details,
including ones from the artifact.  
For example, inspired by a network encountered by one of our team members, 
the CCA switchbox scenario (Figure~\ref{fig:switchbox}) 
describes a corporate network with switchbox.
We present this scenario using prose and a system diagram and use it
to target the concept of identifying security goals.
As revealed in our cognitive interviews, these artifacts inspired and challenged students to apply concepts to complex situations. 
A difficulty in adding artifacts is to maintain focus on important timeless concepts and to minimize emphasizing particular time-limited facts, languages, or conventions.

Responses from our new crowdsourcing experiment (Section~\ref{sec:turk}) suggest that some subjects were confused about how many LANs could be connected through the switch simultaneously.  Consequently, we made one minor clarifying edit to 
the last sentence of the scenario: we changed ``switch that physically connects the computer to the selected LAN'' to ``switch that physically connects the computer to exactly one LAN at a time.''

Comparing the CCI ``lost luggage'' scenario to the CCA ``switchbox'' scenario, one can see that the CCI scenario is simple, requiring few details to be clear. On the other hand, the CCA scenario requires the consideration and analysis of a greater number of facts and properties of the system. Some of these facts, such as ``[the accounting LAN] is not connected to the Internet'' and that ``each desktop computer has one network interface card,'' may have been added in discussion as the problem developers required clarification in their discussion of the scenario. In conjunction with the artifact, the scenario serves to constrain the problem in such a way that the system can be well-understood. 

\subsection{Drafting Stems} 
\label{sec:stems}

Drafting a stem requires careful consideration of several points, in the context of the scenario and alternatives.  Each test item should primarily target one of the five core concepts, though to some degree it might involve additional concepts. The stem should be meaningful by itself, and an expert should be able to answer it even without being provided any of the alternatives.  We try to to keep each stem as focused and short as reasonably possible.  To this end, we try to place most of the details into the scenario, though stems may add a few supplemental details.  Each test item should measure conceptual understanding, not informational knowledge, and not intelligence.  Throughout we consider the ``Vanderbilt'' guidelines~\cite{vanderbilt}, which, among other considerations, caution against negatively worded stems, unless some significant learning outcome depends on such negativity 
(e.g., ``What should you {\bf NOT} use to extinguish an oil fire?'')


There are many pitfalls to avoid, including unclear wording, ambiguity, admitting multiple alternatives, using unfamiliar words, and being too easy or too hard.  As a rule of thumb, to yield useful information, the difficulty of each test item should be set so that at least 10\%, and at most 90\%, of the subjects answer correctly. 
We try hard to leave nothing to the subject's imagination, making it a mistake for subjects to add details or assumptions of their own creation that are not explicitly in the scenario or stem.

To carry out the detailed work of crafting test items, we created a problem development group, whose regular initial members were cybersecurity experts Sherman, Golaszewski, and Scheponik. In fall 2018, Peterson joined the group.  Often during our weekly CATS conference calls, we would present a new CCI item to Herman and Oliva, who are not cybersecurity experts. It was helpful to hear the reactions of someone reading the item for the first time and of someone who knows little about cybersecurity.  An expert in MCQs, Herman was especially helpful in identifying unintentional clues in the item.  Herman and Oliva were less useful in reviewing the more technical CCA test items.

We created and refined stems in a highly iterative process.  Before each meeting, one member of the problem development group would prepare an idea, 
based strongly on one of our core concepts.  
During the meeting, this member would present their suggestion through a shared Google Doc, triggering a lively discussion.  Other members of the group would raise objections and offer possible improvements, while simultaneously editing the shared document.  
Having exactly three or four members present worked extremely well for us,
to provide the diverse perspectives necessary to identify and correct issues, while keeping the discussion controlled enough to avoid anarchy and to permit everyone to engage.  
Over time we became more efficient and skilled at crafting test items,
because we could better overcome predictable difficulties and avoid common missteps.

Sometimes, especially after receiving feedback from students or experts, we would reexamine a previously drafted test item.  Having a fresh look after the passage of several weeks often helped us to see new issues and improvements.  

Continuing the switchbox example from Section~\ref{sec:scenarios},
Figure~\ref{fig:switchbox-stem} gives three versions of this CCA stem
during its evolution.  In Version~1, we deemed the open-ended phrasing as
too subjective since it is impossible for the subject to determine definitely the
network design's primary intent.  This type of open-ended stem risks leading to
multiple acceptable alternatives, or to one obviously correct alternative and four easily rejected distractors. Neither of these outcomes would be acceptable.  

In Version~2, instead of asking about the designer's intent, we ask about security goals that this design supports.  We also settled on a unified language for stems, using the verb ``choose,'' which assertively emphasizes that the subject should select one best answer
from the available choices.
This careful wording permits the possibility that the design might support multiple security goals, while encouraging one of the security goals to be more strongly supported than the others.

\begin{figure}[b]
    \centering
    \begin{tabular}{ll}
        Version 1: &  What security goal is this design primarily intended to meet?\\
        Version 2: &  Choose the security goal that this design best supports.\\
        Version 3: & Choose the action that this design best prevents.
    \end{tabular}
    \caption{Evolution of the stem for the CCA switchbox test item.}
    \label{fig:switchbox-stem}
\end{figure}

Seeking even greater clarity and less possible debate over what is the best answer, we iterated one more time.  In Version~3, we move away from the possibly subjective phrase ``goal that this design best supports'' and instead focus on the more concrete ``action that this design best prevents.''  This new wording also solves another issue: because the given design is poor, we did not wish to encourage subjects to think that we were praising the design.   This example illustrates the lengthy, careful, detailed deliberations we carried out to create and refine stems.

\subsection{Developing Distractors} 
\label{sec:distractors}

Developing effective distractors (incorrect answer choices) is one of the hardest aspects of creating test items.  An expert should be able to select the best answer easily, but a student with poor conceptual understanding should find at least one of the distractors attractive. Whenever possible, we based distractors on misconceptions uncovered during our student interviews~\cite{Thompson2018}.  Concretely doing so was not always possible in part because the interviews did not cover all of the scenarios ultimately created, so we also based distractors on general misconceptions (e.g., encryption solves everything).
The difficulty is to develop enough distractors while satisfying the many
constraints and objectives.

For simplicity, we decided that each test item would have exactly 
one best (but not necessarily ideal) answer.  
To simplify statistical analysis of our assessments~\cite{FIE2019}, 
we decided that each test item would have the same number of alternatives.
To reduce the likelihood of guessing correctly, and to reduce the required number of test items,
we also decided that the number of alternatives would be exactly five.  
There is no compelling requirement to use five;
other teams might choose a different number (e.g., 2--6).  

Usually, it is fairly easy to think of two or three promising distractors.  The main difficulty is coming up with the fourth.  For this reason, test creators might choose to present four rather than five alternatives.  Using only four alternatives (versus five) increases the likelihood of a correct guess; nevertheless, using four alternatives would be fine, provided there are enough test items to yield the desired statistical confidence
in student scores.

As we do when drafting stems, we consider the ``Vanderbilt'' guidelines~\cite{vanderbilt}, which include the following:
All alternatives should be plausible (none should be silly), and each distractor should represent some misconception.
Each alternative should be as short as reasonably possible.  
The alternatives should be mutually exclusive (none should overlap).
The alternatives should be relatively homogeneous (none should stand out as different,
for example, in structure, format, grammar, or length).
If all alternatives share a common word or phrase, that phrase should be moved to the stem.

Care should be taken to avoid leaking clues, both within a test item and between different test items.  In particular, avoid leaking clues with strong diction, length, or any unusual
difference among alternatives.
Never use the alternatives ``all of the above'' or ``none of the above;'' these
alternatives complicate statistical analysis and provide little insight into student understanding of concepts.
If a negative word (e.g., ``{\bf NOT}'') appears in a test item, it should be emphasized
to minimize the chance of student misunderstandings.
As noted in Section~\ref{sec:stems}, typically stems should be worded in a positive way.

To develop distractors, we used the same interactive iterative process described in Section~\ref{sec:stems}.  We would begin with the correct alternative, which for our convenience only during test item development, we always listed as Alternative~A. 
Sometimes we would develop five or six distractors, and later pick the four selected most frequently by students. To overcome issues (e.g., ambiguity, possible multiple best answers, or difficulty coming up with more distractors), we usually added more
details to the scenario or stem.  For example, to constrain the problem, we might clarify the assumptions or adversarial model.

Reflecting on the difficulty of conducting student interviews and brainstorming quality distractors, we investigated alternate ways to develop distractors~\cite{CATS-turk,crowdsourcing_arxiv}.
One way is to have students from the targeted population answer stems {\it without} being offered any alternatives.  By construction, popular incorrect answers 
are distractors that some subjects will find attractive.  
This method has the advantage of using a specific actual stem. 
For some test items, we did so using student responses from our student interviews.
We could not do so for all test items because
we created some of our stems after our interviews.


An even more intriguing variation is to collect such student responses through
crowdsourcing (e.g., using Amazon Mechanical Turk~\cite{turk}). We were able to do
so easily and inexpensively overnight~\cite{crowdsourcing_arxiv,CATS-turk}.  
The main challenges are
the inability to control the worker population adequately, and the high prevalence of cheaters (e.g., electronic bots deployed to collect worker fees, or human workers who do not expend a genuine effort to answer the stem).  Nevertheless, even if the overwhelming majority of responses are gibberish, the process is successful if one can extract at least four attractive distractors.  Regardless, the responses require grouping and refinement.
Using crowdsourcing to generate distractors holds great promise and could be significantly improved with verifiable controls on the desired workers.

\begin{figure}[t]
    \centering
    \begin{tabular}{lll}
    \multicolumn{3}{l}{What security goal is this design primarily intended to meet?}\\
    {\hspace*{.1in}} 
         & A & Prevent employees from accessing accounting data from home.\\
         & B & Prevent the accounting LAN from being infected by malware.\\
         & C & Prevent employees from using wireless connections in Accounting.\\
         & D & Prevent accounting data from being exfiltrated.\\
         & E & Ensure that only authorized users can access the accounting network.\\
         \multicolumn{3}{c}{{\it Initial Version}} \\[12pt]
      \multicolumn{3}{l}{Choose the action that this design best prevents:} \\
         & A & Employees accessing the accounting server from home.\\ 
         & B & Infecting the accounting LAN with malware.\\
         & C & Computer A communicating with Computer B.\\
         & D & Emailing accounting data.\\
         & E & Accessing the accounting network without authorization.\\
       \multicolumn{3}{c}{{\it Final Version}} \\
    \end{tabular}
    \caption{Evolution of the alternatives for the CCA switchbox test item.}
    \label{fig:switchbox-alternatives}
\end{figure}

Continuing the switchbox example from Sections~\ref{sec:scenarios} and~\ref{sec:stems}, 
we explain how we drafted the distractors and how they evolved.  Originally, when we
had created the scenario, we had wanted the correct answer to be preventing data from being exfiltrated from the accounting LAN (Alternative~D is a more specific instance of this idea).
Because the system design does not prevent this action, we settled on the correct answer being preventing access to the accounting LAN.  To make the correct answer less obvious, we worded
it specifically about employees accessing the accounting LAN from home.  
Intentionally, we chose not to use a broader wording about people accessing the accounting LAN from the Internet, which subjects in our new crowdsourcing experiment (Section~\ref{sec:turk}) subsequently came up with and preferred when presented the open-ended stem without any alternatives.

Figure~\ref{fig:switchbox-alternatives} illustrates the evolution of five
alternatives to this CCA stem.  As discussed in Section~\ref{sec:stems}, the initial version of this test item has undesirable ambiguity.  A second issue is that that
Alternative~C is highly implausible because there is nothing in the scenario or stem that involves wireless connections.  A third issue is that the word ``prevent" appears in four alternatives.

The final version of the test item makes three improvements.  
First, we cast the stem and alternatives in a more concrete and less ambiguous fashion.  
Second, Alternative~C appears more plausible.  Third, we moved the word ``prevent'' into the stem. In each version, the best answer is Alternative~A.

Reflecting on data from our new crowdsourcing experiment (Section~\ref{sec:turk}), 
the problem development committee met to revisit the switchbox example again.
Given that significantly more respondents chose Alternative~E over~A, we wondered why and carefully reexamined the relative merits of these two competing alternatives.
Although we still prefer Alternative~A over~E (mainly because E emphasizes 
{\it authorization}, while the switch deals only with physical {\it access}), we recognize that E has some merit, especially with regard to malicious activity originating from the Internet. To make Alternative~A unarguably better than E, we reworded E more narrowly: 
``User of Computer~B from accessing the accounting LAN without authorization.''

We made this change to Alternative~E, not per se because many subjects chose E over A, but because we recognized a minor issue with Alternative~E. The experimental data brought this issue to our attention.  It is common in concept inventories that, for some test items, many subjects will prefer one of the distractors over the correct alternative, 
and this outcome is fine.


\subsection{Generating Educational Materials} 
\label{sec:education}

Pleased with the student engagement stimulated by our scenarios and associated interview prompts, we realized that these scenarios can make effective case studies through which students can explore cybersecurity concepts.  To this end, we published a paper presenting six of our favorite scenarios, together with our exemplary responses and selected misconceptions students revealed reasoning about these scenarios~\cite{exploring2016}.  These scenarios provide an excellent way to learn core concepts in cybersecurity in thought-provoking complex practical contexts.

Although we have not yet done so, it would be possible to create additional learning activities inspired by these scenarios, including structured discussions, design and analysis challenges, and lab exercises.

Also, one could prepare a learning document that consisted of new MCQ test items together with detailed discussions of their answers.  For such
a document, it would be helpful to permit 0--5 correct answers for each test item.  Doing so would reap two benefits:  First, it would simplify creating test items because it would avoid the challenge of requiring exactly one best answer.  Second, it would create an authentic learning moment to be able to discuss a variety of possible answers.  In particular, typically there can be many possible ways to design, attack, and defend a system.

\subsection{Performing Expert Reviews} 
\label{sec:experts}

We obtained feedback from cybersecurity experts on our draft assessments in three ways.  
First, we received informal feedback.
Second, experts reviewed the CCI and CCA during our workshops and hackathons.
Third, as part of our validation studies, we conducted more formal expert reviews~\cite{FIE2019}.
These experts include cybersecurity educators and practitioners from government and private industry.
We recruited the experts through email, web announcements, and at conferences.

Each expert took the CCI or CCA online (initially through SurveyMonkey, and later via PrairieLearn \cite{west_2015}).  For each test item, they first selected their answer choice.
Next, we revealed what we considered to be the correct answer.  We then invited the expert to write comments, and the expert rated the test item on the following scale: accept, accept with minor revisions, accept with major revisions, reject.  Following the online activity, we engaged a group of experts in a discussion about selected test items.

We found this feedback very useful.  It was reassuring to learn that the experts agreed that the test items probe understanding of the targeted concepts and that they agreed with our answer choices.  The experts helped us identify issues with some of the test items.  Because resolving such issues can be very time consuming, we preferred to keep the discussion focused on identifying the issues.  In days following an expert review, the problem development group refined any problematic test items.

Some experts disagreed with a few of our answer choices.  Usually they changed their opinion after hearing our explanations.  Sometimes the disagreement reflected an ambiguity or minor error in the test item, which we later resolved, often by providing more details in the scenario.  Usually, when experts disagreed with our answer choice, it was because they made an extra assumption not stated in the scenario or stem.  We instructed subjects not to make any unstated assumptions, but some common sense assumptions are often required.  
Experts, more so than students, struggled with the
challenge of what to assume, possibly because of their extensive knowledge of 
possibly relevant factors.
Dealing with this challenge is one of the special difficulties in creating cybersecurity concept inventories.

\subsection{Recruiting Test Subjects} 
\label{sec:subjects}

Recruiting large numbers of test subjects turned out to be much more challenging than we had expected. We are validating the CCI in 2018--2020 through small-scale pilot testing (100--200 subjects), cognitive interviews, expert review, and large-scale psychometric testing (1000 subjects). For the CCA, we are following a parallel plan in 2019--2021.  Staggering the validations of the CCI and CCA helps balance our work and permits us to adapt lessons learned from the earlier years. It is difficult to recruit others to invest their scarce time helping advance our project.

For pilot testing of the CCI, we recruited 142 students from six schools.
We also recruited 12 experts and carried out approximately seven cognitive interviews. 
We recruited these students by direct contact with educators we knew who were teaching introductory cybersecurity classes.
The overwhelming majority of experts rated every item as measuring appropriate cybersecurity knowledge. Classical test theory~\cite{Bechger:2003,Hamleton:1993,Jorion2015} showed that the CCI is sufficiently reliable for measuring student understanding of cybersecurity concepts and that 
the CCI may be too difficult as a whole~\cite{FIE2019}.
In response to these inputs, we revised the CCI and moved one of the harder test items to the CCA.

In fall 2019, we started to recruit 1000 students to take the CCI so that we can evaluate its quality using item response theory~\cite{Bechger:2003,Hamleton:1993,Jorion2015}, which requires many more subjects than does classical test theory.  Recruiting subjects proved difficult and by late December just under 200 students had enrolled. We hope to recruit another 800 subjects by late spring 2020.  Because of the ongoing COVID-19 pandemic, all of these subjects will take the CCI online.  

By far, the most effective recruitment method has been direct contact with educators (or their associates) we personally know who teach introductory cybersecurity classes.  Sending email to people we do not personally know has been extremely ineffective, with a response rate of approximately~1\%.
Making personal contact at conferences usually resulted in a stated willingness to participate but without subsequent action.  Following up on such promises sometimes increased our yield.
We advertised through email, web, and conferences.
Posting notices in newsletters (e.g., for CyberWatch), and making announcements at PI meetings seemed to have some positive effect.  
We targeted specific likely groups, including people connected with 
Scholarship for Service (SFS) programs and National Centers of Excellence in Cyber Defense Education (CAE).  We also asked participating instructors for referrals to additional instructors.
At conferences, passing out slips of paper with URLs and QR-codes did not work well.
Keeping our recruitment pitch brief, limited, and simple helped.

For instructors who enroll in our validation study,  some administer the CCI in class through our web-based system; others suggest it as an optional out-of-class activity.  For the latter group, offering some type of incentive (e.g., extra credit) has been critical.  It takes students approximately one hour to complete the CCI and two hours to take the CCA.  Virtually no students took the CCI in classes where the instructor offered no incentive.

Timing has been another issue.  Some colleges and universities offer an introductory cybersecurity course only once a year, and students are not ready to take the assessment until towards the end of the course.  Furthermore, contacting students after they have completed the course has been mostly ineffective.  It has improved our yield to ask an instructor when they will be able to administer the assessment, and then follow up when the time comes closer.

Recruiting 1000 subjects for the CCA might be even harder because the target population is smaller.

\subsection{Organizing and Running Hackathons} 
\label{sec:hackathon}

In February 2018, we hosted a two-day ``Hackathon'' for 17 cybersecurity educators and professionals from across the nation to generate multiple-choice test items for the CCA, and to refine draft items for the CCI and CCA~\cite{hackathon}.  To collect additional expert feedback, we also held shorter workshops at the 2019 National Cyber summit and 2019 USENIX Security Symposium. We had to cancel another workshop planned for June 2020, to have been coordinated with the Colloquium for Information System Security (CISSE), 
due to the COVID-19 pandemic. 

Our focused Hackathon engaged experts, generated useful inputs, and promoted awareness of the project.  Each participant focused on one of the following tasks: 1)~generating new scenarios and stems; 2)~extending CCI items into CCA items, and generating new answer choices for new scenarios and stems from Task~1; and 3)~reviewing and refining draft CCA test items. These tasks kept each team fully engaged throughout the Hackathon. Each participant chose what team to join, based in part on their skill sets. The event took place at an off-campus conference center, two days before the ACM Special Interest Group on Computer Science Education (SIGSCE) conference in Baltimore, Maryland. Thirteen experts came from universities and two each came from industry and government, respectively. Participants took the CCI at the beginning of the first day and the CCA at the beginning of the second day.  We paid participant costs from our NSF grant.  

The most difficult challenge in organizing our Hackathon was obtaining commitments of participants to attend.  We had originally planned to hold the event in October 2017, but rescheduled due to low interest.  Holding the event immediately before a major computer science education conference helped. As for recruiting instructors in our pilot studies, direct one-on-one person contact was our most effective communication method.

With the hopes of more easily attracting participants, for the shorter workshops we tried a slightly different strategy.  We coordinated with conference organizers to hold a four-hour workshop at the conference site, typically immediately before the main conference began.  Although recruitment remained a challenge and the shorter workshop provided fewer inputs, this strategy mostly worked better.  To our surprise, offering food at these shorter workshops seemed to make no difference.

Initially, we had thought that offering to pay participant expenses (airfare, hotel, per-diem) would help motivate participants to attend.  This strategy, however, encountered two issues.
First, many people already had funding and they could not absorb additional reimbursements.
Second, reimbursements require significant paperwork and administrative procedures, which are inconvenient for participants and organizers.  When possible, it is simpler and more effective instead to pay honoraria.

\subsection{Building Community Acceptance} 
\label{sec:acceptance}

Many concept inventories remain largely unused, poorly accepted, and poorly adopted.  
Members of the subject community do not recognize the importance of the concepts they assess or the effectiveness of the assessments. To avoid this fate, great care must be taken to build acceptance throughout the creation and validation process.

The CATS Project is pursuing three strategies to build widespread acceptance. First, we are following a well-accepted scientific methodology for creating and validating the CCI and CCA.
Second, throughout the project we are engaging experts, including in our Delphi processes that identified core concepts, in our workshops and hackathons that develop and refine test items, and
in our experts reviews of draft assessments.
Third, we are presenting our work at conferences to seek feedback and promote awareness of our
assessment tools.

It remains to be seen how widely the CCI and CCA will be used.  One special characteristic that may help is the driving recognition of the strong need to prepare more cybersecurity practitioners.
It would probably further advance our assessments to carry out an even more active program of presenting our work to a wide range of cybersecurity educators.
We hope that many researchers will use the CCI and CCA to identify and measure
effective approaches to teaching and learning cybersecurity.

\subsection{Forming a Team and Nurturing Collaboration} 
\label{sec:collaboration}

With experts in cybersecurity, engineering education, and educational assessment, our team covers all of the key areas needed to accomplish our goals.  In particular, Herman brings prior experience creating, validating, and using concept inventories.  A postdoc, several graduate students, an undergraduate student, and a high school student also add useful diversity, talent, and labor.  For example, one PhD student brings practical cybersecurity experience from NSA and private industry. This interdisciplinary, inter-institution collaboration including a variety of perspectives has been crucial to our success.

The entire team meets every Thursday on Skype for up to an hour, whenever there is work to be done; we skip meetings if there is nothing to discuss. These meetings help us make strategic decisions and keep us focused on what needs to be done.  In addition, subgroups---such as the problem development group---meet weekly as needed. Oliva assumed primary responsibility for handling all IRB approvals
and conducting student interviews.  Sherman and Herman took primary responsibility for writing all grant proposals and grant reports.

When developing a test item, the problem development group might spend time discussing one or more of the following steps:  high-level brainstorming, creating a scenario, drafting a stem, developing distractors, refining test items, and recording meta-data.  A typical session lasted about two hours or slightly longer.  After initially meeting in person at UMBC, we ultimately preferred to meet remotely while simultaneously editing a Google Doc, which was more efficient and accommodated new team member Peterson from Duluth, MN.  We felt productive if we could make substantial progress on at least one test item per meeting.


When writing a paper, one team member would serve as the primary writer/editor. We delegated tasks such as writing particular sections, preparing figures, programming, and dealing with references.  


Our team functions smoothly and has avoided any major fracas.  Team members have mutual respect for each other and value each person's unique contributions. We resolve conflict in a civil and constructive way. We include everyone in deliberations about all important matters, encouraging frank critical discussions regardless of rank.  This style of decision making helps us make wise choices.

\subsection{Adopting Tools} 
\label{sec:tools}

We used a variety of tools to support various aspects of the project, including tools for conferencing (Skype), document sharing (Google Drive, Github, Overleaf), simultaneous editing (Google Docs, Overleaf), surveys (SurveyMonkey), document preparation ({\LaTeX} with Overleaf, Microsoft Word), data management (Excel), and delivery and analysis of tests (PrairieLearn). We are not familiar with any useful dedicated tool for supporting the development of MCQs.  To avoid the trouble of making and maintaining our own tool, we used a collection of existing tools.  Over time, as we worked on different tasks and experimented with a variety of approaches for interacting with each other, our operational methods and choice of supporting tools evolved.
Three critical tasks presented especially important challenges: 
conferencing to create and refine test items, delivery and analysis of tests, and 
management of test items throughout their life cycle.

Initially, we brainstormed test items at UMBC sitting around a conference table, while one team member wrote notes into an electronic document that we projected onto a screen.  Eventually we found this method of operations suboptimal for several reasons.  First, it was difficult to schedule a time when everyone in the problem-development group could attend. Second, it was inefficient for people to drive to UMBC for such meetings, especially because one member lived in DC and another moved to Richmond.  Third, having only one person edit was inefficient, especially when making simple edits.
Fourth, as often happens by human nature, we tended to waste more time when meeting in person
than meeting online.  Work proceeded more smoothly and efficiently when we switched to online conferencing, during which each participant simultaneously viewed and edited a Google Doc.  Participants could see displays more clearly when looking at their own monitor than looking at a conference room projection. For groups that meet in person, we recommend that each participant bring a laptop to enable simultaneous editing.

When we started our validation studies, we needed a way for students and experts to take our assessments and for experts to comment on them.  After experimenting briefly with SurveyMonkey, we chose PrairieLearn~\cite{west_2015}, 
a tool developed at the University of Illinois, and already used by Herman.  This tool allows subjects to take tests online and provides statistical support for their analysis.  It also supports a number of useful test delivery features, including randomizing answer choices.  PrairieLearn, however, does not support the life cycle of test items.
Also, in PrairieLearn it is inconvenient to enter test items with mathematical expressions
(we awkwardly did so using HTML).

The CATS Project could have benefited greatly from an integrated tool to support the entire life cycle of test items, maintaining the authoritative version of each test item and associated meta-data (comments, difficulty, topics, concepts, and validation statistics), and avoiding any need to copy or translate test items from one system to another (which can introduce errors).  Such a system should also support simultaneous editing, and it should display test items as they will appear to students.  While developing more than 60 test items over several years, it was essential to save detailed notes about each test item in an orderly fashion.

Though PrairieLearn has many strengths, PrairieLearn falls short:  with test items written in HTML, it was difficult to edit the source simultaneously.  We tried using the Google-recommended HTML editor Edity but found it cumbersome, due in part to
synchronization difficulties across multiple editors.
We ended up making most edits in a Google Doc and 
then manually translating the results into PrairieLearn.  Curiously, PrairieLearn provides no automatic way to generate a PDF file of a test, so we wrote a script to do so semi-automatically, by converting the PrairieLearn HTML to {\LaTeX} using pandoc.
Although we can define assessments in PrairieLearn, the system provides no
high-level support for helping us to decide which test items (from a larger bank of items)
should be included to achieve our desired goal of having 
five test items of various difficulty levels from each of the five targeted concepts.

Initially we collaborated writing documents using Github
but eventually found Overleaf much easier to use.  
We prepared most of our publications using {\LaTeX}.  
A few times we experimented with Microsoft Word, 
each time regretting our choice (due in part to poor support of mathematics, simultaneous editing, and fine-grain document control).  
We found Overleaf very convenient to use because it supports simultaneous editing and it provides a uniform compilation environment. 

Because long email threads can be confusing,
we are considering deploying Slack to manage channels 
of text messages that persist beyond the live exchanges.
Such channels would be useful for many tasks of the project, including
writing papers.

\subsection{Obtaining and Using Funding} 
\label{sec:funding}

Three grants (each collaborative between UMBC and Illinois) have directly funded the CATS project: two one-year CAE-R grants from NSA, and one three-year SFS capacity grant from NSF.  In addition, UMBC's main SFS grant provided additional support.  Funds have supported a post-doc, graduate RAs, faculty summer support, travel, and workshops.  This funding has been helpful in promoting collaborations, including 
between our two institutions, within each institution, and more broadly through workshops.  Especially initially, the prospects of possible funding helped Sherman and Herman forge a new partnership.  

Some of the factors that likely contributed to our successful funding include the following: We have a collaborating team that covers the needed areas of expertise. We submitted convincing detailed research plans (for example, we secured agreements with most of the Delphi experts by proposal time).  At each step, we presented preliminary accomplishments from the previous steps. NSF recognized the value of a concept inventory that could be used as a scientific instrument for measuring the effectiveness of various approaches to cybersecurity education (see Section~\ref{sec:genesis}).

The CATS Project has produced much more than MCQs, and it has created more MCQs than the 50
on the CCI and CCA.
For example, the CATS Project has published eight research papers, including on
core concepts of cybersecurity~\cite{Delphi2016},
student misconceptions about cybersecurity~\cite{Thompson2018}, 
using crowdsourcing to generate distractors~\cite{crowdsourcing_arxiv,CATS-turk}, and
case studies for teaching cybersecurity~\cite{exploring2016}.
Keeping in mind that most of our funding has supported research activities beyond creating MCQs, it is nevertheless interesting to try to approximate the cost of creating validated test items. Dividing the project's total combined funding by the 50 MCQs on the CCI and CCA yields a cost of \$21,756 per test item.
There is commercial potential for creating cybersecurity assessment tools, but any financially successful company would  have to develop efficient processes and amortize its expenses over a large number
of test takers.
\section{Discussion}
\label{sec:discussion}

Working on the CATS Project we have learned a lot about creating and validating 
MCQ concept inventories.  We now discuss some of our most notable takeaways from our
evolving relationship with MCQs.  In particular, we discuss
the difficulty of creating effective MCQs, 
the special challenges of cybersecurity MCQs,
the value of using scenarios in test items,
the advantages and limitations of MCQs, and
some general advice for creating test items.

Creating MCQs that effectively reveal mastery of core concepts is difficult and time consuming. Although we have become more efficient, 
it still takes us many hours to conceive of, draft, and refine a test item.  Great care is needed to ensure that 
the test item is clear to all, there is exactly one best answer from among five plausible alternatives, 
students who understand the underlying concept will be able to select the correct answer, and
students who do not understand the underlying concept will find some of the distractors appealing.
Test items should not be informational or measures of intelligence.
Care must be taken to minimize the chance that
some students might become confused for unintended reasons
that do not necessarily relate to conceptual understanding,
such as about a particular detail or word choice.  

Cybersecurity presents some special difficulties for creating and answering MCQs.
First, details often matter greatly. In cybersecurity, issues are frequently subtle, and the best answer often hinges on the details and adversarial model. 
It is important to provide enough details (but not too many), and 
doing so is challenging within the constraints of a MCQ.
Second, it is often difficult to ensure that there is exactly one best answer.
For example, typically there are many potential designs and vulnerabilities.
An ``ideal'' answer, if it existed, might appear too obvious.
To deal with this difficulty, we usually add details to the scenario and stem; 
we emphasize that the task is to select the best alternative 
(which might not be a perfect answer) from the available choices; 
and we sometimes frame the question in terms of comparing alternatives (e.g., best or worst).
Third, it is problematic to create MCQs questions based on attractive open-ended questions, such as, ``Discuss potential vulnerabilities of this system.''
Often it can be much more challenging to think of a vulnerability than to recognize it
when stated as an alternative.  Listing a clever vulnerability or attack as an answer choice
can spoil an otherwise beautiful question.  Again, to deal with this difficulty,
it can be helpful to ask students to compare alternatives.
These difficulties are common to other domains (e.g., engineering design), and they are strongly
present in cybersecurity.

We found it helpful to create test items that 
comprise a scenario, stem, and alternatives. 
Each scenario describes an engaging situation in sufficient detail to motivate a meaningful cybersecurity challenge.  For the CCA, to add technical substance, we typically included some artifact, such as a
protocol, system design, log file, or source code.
Our strategy was to put most of the details into the scenario---rather than into the stem---so that the stem could be as short and straightforward as reasonably possible.  This strategy worked well.
We created an initial set of twelve scenarios for our talk-aloud interviews to uncover student misconceptions~\cite{Thompson2018}.
The scenarios provide useful building blocks for other educational activities, including
exercises and case studies~\cite{exploring2016}.
In creating scenarios, it is necessary to balance depth, breadth, richness, and length.

Originally, we had expected to be able easily to create several different stems for each scenario, which should reduce the time needed to complete the assessment and reduce the cognitive load required to read and process details. For example, in the {\it Scholastic Assessment Test (SAT)}, several test items share a common
reading passage. Although several of our test items share a common scenario, unexpectedly we found it problematic to do so for many scenarios. The reason is that, for each stem, we felt the need to add many details to the scenario to ensure that there would be exactly one best answer, and such details were often specific to the stem.
After customizing a scenario in this way, the scenario often lost the generality needed to work for other stems.  Whenever we do share a scenario, for clarity, we repeat it verbatim.

In keeping with the predominant format of CIs, 
we chose to use MCQs because
they are relatively fast and easy to administer and grade;
there is a well established theory for validating them and interpreting their results~\cite{Pellegrino2011}; 
it is possible to create effective MCQs; 
and they are relatively easier to create than are 
some alternatives (e.g., computer simulations).
Albeit more complex and expensive, alternatives to MCQs may offer ways to overcome some of the limitations of MCQs.  These alternatives include
open-ended design and analysis tasks,
hands-on challenges,
games, 
and interactive simulations.
We leave as open problems to create and validate cybersecurity assessment tools based on alternatives to MCQs.

To generate effective MCQ test items, we found it useful to work in a diverse collaborating team,
to draw from our life experiences, to engage in an iterative development process, and
to be familiar with the principles and craft of drafting questions
and the special challenges of writing multiple-choice questions about security scenarios.
It is helpful to become
familiar with a variety of question types (e.g., identifying the best or worst),
as well as question themes (e.g., design, attack, or defend).  
When faced with a difficult challenge, such as reducing ambiguity or ensuring a single best answer, 
we found the most useful strategy to be adding details to the scenario and stem, to clarify
assumptions and the adversarial model.
Introducing artifacts is an engaging way to deepen the required technical knowledge.
To promote consistency, we maintained a style guide for notation, format, and spelling.



\section{Conclusion}
\label{sec:conclusion}

When we started the CATS Project six years ago, we would have loved to have been able to read a paper
documenting experiences and lessons learned from creating and validating a concept inventory.
The absence of such a paper motivated us to write this one.
As we addressed each key step of the process creating two cybersecurity concept inventories,
we selected an approach that seemed appropriate for our situation, adjusting our approach as needed.

One of the most important factors contributing to our success is collaboration.
Our diverse team includes experts in cybersecurity, systems, engineering education, and educational assessment.
When creating and refining test items, it is very helpful to have inputs from a variety of perspectives.
Often in security, a small observation makes a significant difference.  Openness and a willingness to
be self-critical helps our team stay objectively focused on the project goals and tasks at hand.

When we started the project, many of us had a very poor opinion of MCQs, based primarily on the weak examples we had seen throughout our lives.  Many MCQs are informational, thoughtless, and flawed in ways that can permit answering questions correctly without knowledge of the subject.\footnote{As an experiment, select answers to your favorite cybersecurity certification exam looking only at the answer choices and not at the question stems.  If you can score significantly better than random guessing, the exam is defective.}  Over time, we came to realize that it is possible to create excellent conceptual MCQs, though doing so is difficult and time consuming.

We believe strongly that the core of cybersecurity is {\it adversarial thinking}---managing information and trust in an adversarial cyber world, in which there are malicious adversaries who 
aim to carry out their nefarious goals and defeat the objectives of others. 
Appropriately, our assessment tools focus on five core concepts of adversarial thinking,
identified in our Delphi studies.

Our work on the CATS Project continues as we complete the validation and refinement of the CCI and CCA
with expert review, cognitive interviews, small-scale pilot testing, and large-scale
psychometric testing.  We invite and welcome you to participate in these 
steps.\footnote{Contact Alan Sherman (sherman@umbc.edu).}
Following these validations, we plan to apply these assessment tools to measure the
effectiveness of various approaches to teaching and learning cybersecurity.


We hope that others may benefit from our experiences with the CATS Project.


\bigskip \noindent {\bf Acknowledgments.}
We thank the many people who contributed to the CATS project as
Delphi experts, interview subjects, Hackathon participants, 
expert reviewers, student subjects, and 
former team members, including 
Michael Neary,
Spencer Offenberger,
Geet Parekh,
Konstantinos Patsourakos,
Dhananjay Phatak, and
Julia Thompson. 
Support for this research was provided in part 
by the U.S. Department of Defense under CAE-R grants 
H98230-15-1-0294, H98230-15-1-0273, H98230-17-1-0349, H98230-17-1-0347; 
and by the National Science Foundation under 
UMBC SFS grants DGE-1241576, 1753681, 
and SFS Capacity Grants DGE-1819521, 1820531.


\bibliography{CATS-bib}
\bibliographystyle{plain}  

\bigskip \noindent 
{\it Invited paper to be presented June 2--4 at the 2020 National Cyber Summit in Huntsville, AL.\\(April 9, 2020)}


\clearpage
\appendix
\section{A Crowdsourcing Experiment}
\label{sec:turk}

 
 To investigate how subjects respond to the initial and final versions of the CCA switchbox question (see Sections~\ref{sec:scenarios}--\ref{sec:distractors}), we polled 100 workers on Amazon Mechanical Turk (AMT)~\cite{turk}.  We also explored the strategy of generating distractors through crowdsourcing, continuing a previous research idea of ours~\cite{crowdsourcing_arxiv,CATS-turk}. Specifically, we sought responses from another 200 workers to  the initial and final stems without providing any alternatives
 
 On March 29--30, 2020, at separate times, we posted four separate tasks on AMT. Tasks 1 and 2 presented the CCA switchbox test item with alternatives, for the initial and final versions of the test item, respectively.
 Tasks~3 and~4 presented the CCA switchbox stem with {\it no} alternatives, for the initial and final versions of the stem, respectively.
 
 For Tasks~1--2, we solicited 50 workers each, and for Tasks~3--4, we solicited 100 workers each. For each task we sought human workers 
 who graduated from college with a major in computer science or related field. We offered a reward of \$0.25 per completed valid response and set a time limit of 10 minutes per task.
 
\begin{figure}[b] 
    \centering
    \begin{tabular}{lll}
    \multicolumn{3}{l}{Question 1: What is eleven plus {\includegraphics[height=10pt]{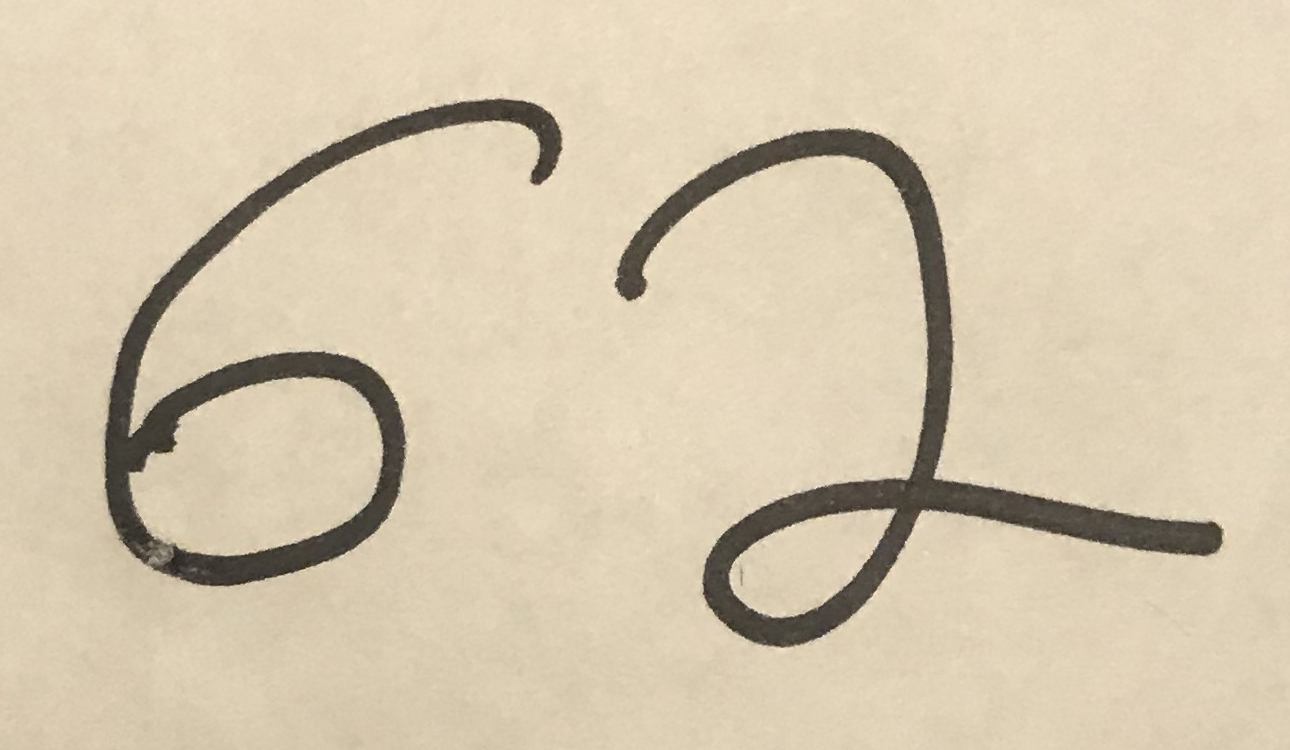}}?
    Express your answer in binary.}\\[12pt]
      \multicolumn{3}{l}{Question 2:  In what major did you graduate from college?} \\
         & A & English\\         
         & B & History\\
         & C & Chemistry\\
         & D & Computer science or related field\\
         & E & Business\\  
         & F & I did not graduate from college\\
    \end{tabular}
    \caption{Control questions for the crowdsourcing experiment on AMT.
    Question~1 aims to exclude bots and subjects who do not know any computer science.}
    \label{fig:control}
\end{figure}

 Expecting computer bots and human cheaters (people who do not make a genuine effort to answer the question), we included two control questions (see Figure~\ref{fig:control}), in addition to the main test item. Because we expected many humans to lie about their college major, we constructed the first control question to detect subjects who were not human or who knew nothing about computer science. All computer science majors should be very familiar with the binary number system. We expect that most bots will be unable to perform the visual, linguistic, and cognitive processing required to answer Question~1. 
 Because $11+62=73=64+8+1$, the answer to Question~1 is $1001001$.
 
 We received a total of 547 responses, of which we deemed 425 (78\%) valid.  
 As often happens with AMT, we received {\it more} responses than we
 had solicited, because some workers responded directly without payment to our
 SurveyMonkey form, bypassing AMT.
 More specifically, we received  40, 108, 194, 205 responses for Tasks~1--4, respectively,
 for which 40 (100\%), 108 (100\%), 120 (62\%), 157 (77\%)
 were valid, respectively.
 In this filtering, we considered a response valid if and only if it was non-blank and appeared to answer the question in a meaningful way.  We excluded responses that did not pertain to the subject matter (e.g., song lyrics), or that did not appear to reflect genuine effort (e.g., ``OPEN ENDED RESPONSE'').
 
 Only ten (3\%) of the valid responses included a correct answer to Question~1. Only 27 (7\%) of the workers with valid responses stated that they majored in computer science or related field, of whom only two (7\%) answered Question~1 correctly. Thus, the workers who responded to our tasks are very different from the intended population for the CCA.

 Originally, we had planned to deem a response valid if and only if
 it included answers to all three questions, with correct answers to each of the two control questions.  Instead, we continued to analyze our data adopting the extremely lenient definition described above, understanding that the results would not be relevant to the CCA.
 
\begin{figure}[t] 
    \centering
    \includegraphics[width=0.75\textwidth,scale=1.5]{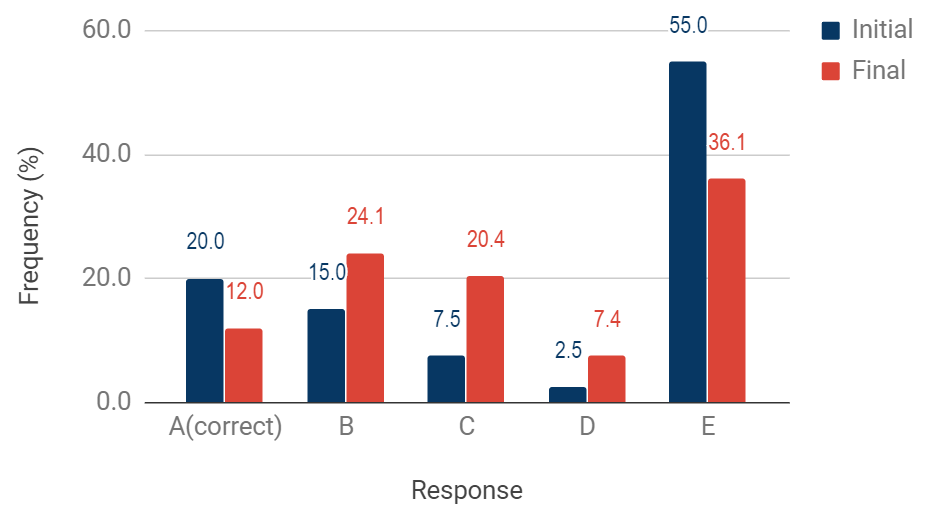}
    \caption{Histogram of responses to the CCA switchbox test item, from AMT workers, for the initial and final versions of the test item. There were 148 valid responses total, 40 for the initial version, and 108 for the final version.} 
    \label{fig:hist12}
\end{figure} 

We processed results from Tasks 3--4 as follows, separately for each task.
 First, we discarded the invalid responses.  
 Second, we identified which valid responses matched existing alternatives.  
 Third, we grouped the remaining (non-matching) valid new responses into equivalency classes. Fourth, we refined a canonical response for each of these
 equivalency classes (see Figure~\ref{fig:new-distractors}).  The most time-consuming steps were filtering the responses for validity and refining the canonical responses.
 
 Especially considering that the AMT worker population differs from the target CCA audience, 
 it is important to recognize that the best distractors for the CCA are not necessarily the most popular responses from AMT workers.  For this reason, we examined
 all responses.  Figure~\ref{fig:new-distractors} includes two examples, 
 $t*$ for initial version (0.7\%) and $t*$ for final version (1.2\%),
 of interesting {\it unpopular} responses.
 Also, the alternatives should satisfy various constraints, including being
 non-overlapping (see Section~\ref{sec:distractors}), so one should not simply automatically 
 choose the four most popular distractors.

\begin{figure} 
    \centering
    \includegraphics[width=0.75\textwidth,scale=1.5]{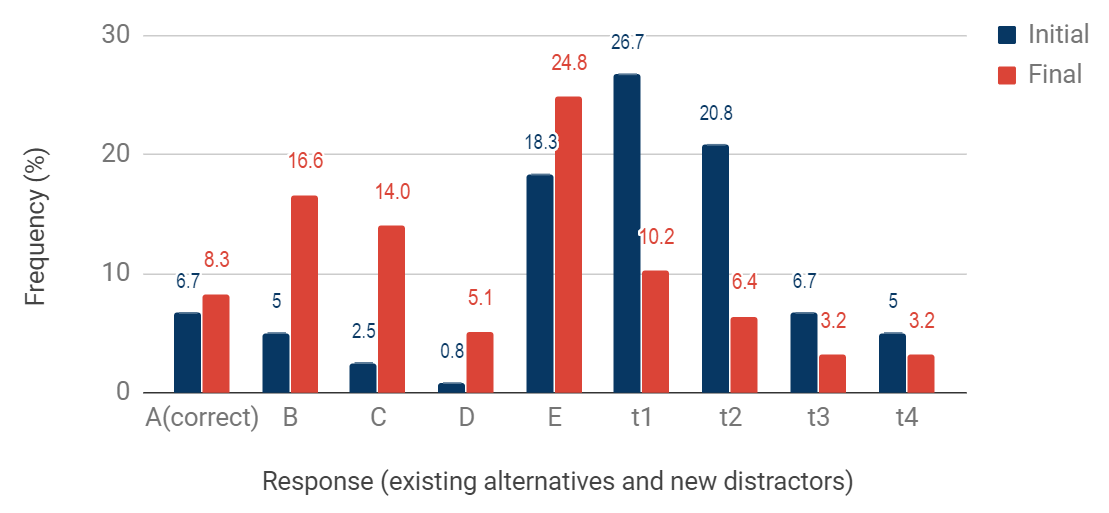}
    \caption{Histogram for equivalency classes of
    selected open-ended responses generated from the CCA switchbox stem.
    These responses are from AMT workers, for the initial and final versions of the stem, presented without any alternatives.  There were 120 valid responses for the initial version, and 157 for the final version.
    A--E are the original alternatives, and $t1$--$t4$ are the four most frequent new responses generated by the workers.  These responses include two alternate phrasings of the correct answer: $t2$ for the initial version, and $t1$ for the final version.
    Percents are with respect to all valid responses and hence do not add up to 100\%.} 
    \label{fig:hist12a}
\end{figure}

\begin{figure} 
    \centering
    \includegraphics[width=0.75\textwidth,scale=1.5]{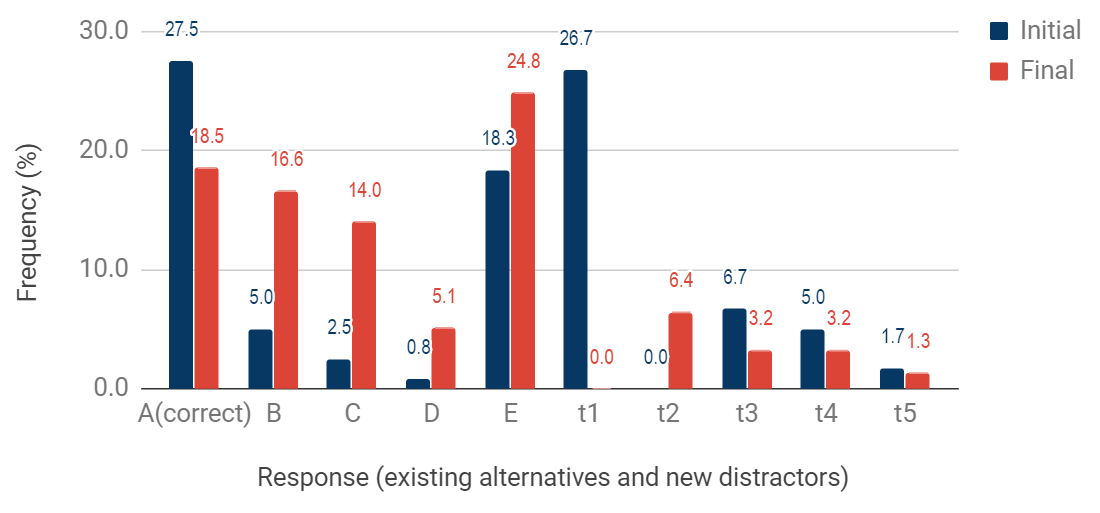} 
    \caption{Histogram for equivalency classes of
    selected open-ended responses generated from the CCA switchbox stem.
    These responses are from AMT workers, for the initial and final versions of the stem, presented without any alternatives.  There were 120 valid responses for the initial version, and 157 for the final version.
    A--E are the original alternatives, and $t1$--$t5$ are the four most frequent new distractors generated by the workers.  The alternate correct responses 
    $t2$ (initial version) and $t1$ (final version) are grouped with Alternative~A.
    Percents are with respect to all valid responses and hence do not add up to 100\%.} 
    \label{fig:hist12b}
\end{figure}

 Figures~\ref{fig:hist12}--\ref{fig:new-distractors} summarize our findings.
 Figure~\ref{fig:hist12} does not reveal any dramatic change in responses between the initial and final versions of the test item.  It is striking how strongly the workers prefer Distractor~E over the correct choice~A.
 Perhaps the workers, who do not know network security, find Distractor~E 
 the most understandable and logical choice.
 Also, its broad wording may be appealing.
 In Section~\ref{sec:distractors}, we explain how we reworded Distractor~E
 to ensure that Alternative~A is now unarguably better than~E.
 
 Figure~\ref{fig:hist12a} shows the popularity of worker-generated responses to the stem, when workers were {\it not} given any alternatives. 
 After making Figure~\ref{fig:hist12a}, we realized it contains a mistake:
 responses $t2$ (initial version) and $t1$ (final version) are {\it correct} answers,
 which should have been matched and grouped with Alternative~A.  We nevertheless include this figure because it is informative.
 Especially for the initial version of the stem, it is notable how more popular the two worker-generated distractors $t2$ (initial version) and $t1$ (final version) are than our distractors. This finding is not unexpected, because, in the final version, we had intentionally chosen Alternative~A over $t1$, to make the correct answer less obvious (see Section~\ref{sec:distractors}).  These data support our belief that
 subjects would find $t1$ more popular than Alternative~A.
 
 Figure~\ref{fig:hist12b} is the corrected version of Figure~\ref{fig:hist12a}, with
 responses $t2$ (initial version) and $t1$ (final version) grouped with the correct answer~A.
 Although new distractor $t1$ (initial version) was very popular, its broad nebulous form
 makes it unlikely to contribute useful information about student conceptual understanding. For the first version of the stem, we identified seven equivalency classes
 of new distractors (excluding the new alternate correct answers);
 for the final version we identified~12.
 
 Because the population of these workers differs greatly from the intended audience for the CCA, these results should not be used to make any inferences about the switchbox test item for the CCA's intended audience.  Nevertheless, our experiment illustrates some notable facts about crowdsourcing with AMT.  (1)~Collecting data is fast and inexpensive:  
 we collected all of our responses within 24 hours paying a total of less than
 \$40 in worker fees (we did not pay for any invalid responses).  (2)~We had virtually no control over the selection of workers, and almost all of them seemed ill-suited for the task (did not answer Question~1 correctly).  Nevertheless, several of the open-ended responses reflected a thoughtful understanding of the scenario.
 (3)~Tasks~3--4 did produce distractors (new and old) of note for the worker population, thereby illustrating the potential of using crowdsourcing to generate distractors, if the selected population could be adequately controlled.
 (4)~Even when the worker population differs from the desired target population, 
 their responses can be useful if they inspire test developers 
 to improve test items and to think of effective distractors.
 
 Despite our disappointment with workers answering Question~1 incorrectly, 
 the experience helped us refine the switchbox test item.
 Reflecting on the data, the problem development team met and made some
 improvements to the scenario and distractors (see discussions near the ends
 of Sections~\ref{sec:scenarios} and~\ref{sec:distractors}).
 Even though the AMT workers represent a different population than our CCA target audience, 
 their responses helped direct our attention to potential ways to improve
 the test item.
 
 We strongly believe in the potential of using crowdsourcing to 
 help generate distractors and improve test items. 
 Being able to verify the credentials of workers assuredly (e.g., cryptographically) 
 would greatly enhance the value of AMT.
 
\begin{figure} 
    \centering
    \begin{tabular}{lll}
    \multicolumn{3}{l}{What security goal is this design primarily intended to meet?}\\
    {\hspace*{.1in}} 
         & A & Prevent employees from accessing accounting data from home. (correct)\\
         & B & Prevent the accounting LAN from being infected by malware.\\
         & C & Prevent employees from using wireless connections in Accounting.\\
         & D & Prevent accounting data from being exfiltrated.\\
         & E & Ensure that only authorized users can access the accounting network.\\[6pt]
         & t1 & Ensure all segments of the network remain independent.\\ 
         & t2 & Keep the accounting server separated from the Internet. (also correct)\\
         & t3 & Prevent malware from affecting the network.\\ 
         & t4 & Prevent the email server from being compromised.\\
         & t5 & Ensure the security of computer B.\\
         & t* & Protect accounting server from DDoS attacks.\\
         \multicolumn{3}{c}{{\it Initial Version}} \\[12pt]
      \multicolumn{3}{l}{Choose the action that this design best prevents:} \\
         & A & Employees accessing the accounting server from home. (correct)\\ 
         & B & Infecting the accounting LAN with malware.\\
         & C & Computer A communicating with Computer B.\\
         & D & Emailing accounting data.\\
         & E & Accessing the accounting network without authorization.\\[6pt]
         & t1 & Accounting server from being accessed through the Internet. (also correct) \\
         & t2 & A computer from accessing both Local Area Networks.\\
         & t3 & Accessing the LAN via the Network Interface Card. \\
         & t4 & Email server from being compromised.\\
         & t5 & Computer C from accessing the Internet.\\
         & t* & Unauthorized access from computer A to server. \\
       \multicolumn{3}{c}{{\it Final Version}} \\ 
    \end{tabular}
    \caption{Existing alternatives (A--E), and refinements of
    the five most frequent new responses ($t1$--$t5$)
    generated for the CCA switchbox stem, from AMT workers, for the initial and final versions of the stem.  There were 120 valid responses for the initial version, and 157 for the final version. Distractors 
    $t*$ for initial version (0.7\%) and $t*$ for final version (1.2\%) 
    are examples of interesting unpopular responses.
    Percents are with respect to all valid responses.}
    \label{fig:new-distractors}
\end{figure}

 
 
 
 
 
 
 


\end{document}